# Studying the Emerging Global Brain:
# Analyzing and Visualizing the Impact of Co-Authorship Teams

**Katy Börner***
Indiana University, SLIS
10th Street & Jordan Avenue
Main Library 019
Bloomington, IN 47405, USA
Phone: (812) 855-3256  Fax: -6166
E-mail: katy@indiana.edu
WWW: http://ella.slis.indiana.edu/~katy

**Luca Dall'Asta**
Laboratoire de Physique Theorique
Université de Paris-Sud, 91405 Orsay, France
E-mail : Luca.Dallasta@th.u-psud.fr

**Weimao Ke**
Indiana University, SLIS
10th Street & Jordan Avenue
Bloomington, IN 47405, USA
E-mail: wke@indiana.edu

**Alessandro Vespignani**
School of Informatics & Biocomplexity Center
Indiana University, Eigenmann Hall, 1900 E. 10th St.
Bloomington, IN 47406, USA
E-mail: alexv@indiana.edu

* To whom all correspondence and proofs are to be addressed.

**Keywords**
Weighted network analysis, co-author networks, citation analysis, information visualization

**Abstract**
This paper introduces a suite of approaches and measures to study the impact of co-authorship teams based on the number of publications and their citations on a local and global scale. In particular, we present a novel weighted graph representation that encodes coupled author-paper networks as a weighted co-authorship graph. This weighted graph representation is applied to a dataset that captures the emergence of a new field of science and comprises 614 papers published by 1,036 unique authors between 1974 and 2004. In order to characterize the properties and evolution of this field we first use four different measures of centrality to identify the impact of authors. A global statistical analysis is performed to characterize the distribution of paper production and paper citations and its correlation with the co-authorship team size. The size of co-authorship clusters over time is examined. Finally, a novel local, author-centered measure based on entropy is applied to determine the global evolution of the field and the identification of the contribution of a single author's impact across all of its co-authorship relations. A visualization of the growth of the weighted co-author network and the results obtained from the statistical analysis indicate a drift towards a more cooperative, global collaboration process as the main drive in the production of scientific knowledge.

**Number of text pages: 12**
**Number of figures: 9**
**Number of tables: 2**

**Summary of results for the nonspecialist:**
The increasing specialization of researchers and practitioners, the accelerated speed of scientific progress and the need to collaborate across institutional and disciplinary boundaries is fostering the formation of high-impact co-authorship teams. This paper presents novel methods to analyze and visualize the size, interconnections and impact of co-author teams over time. The methods were applied to examine the evolution of a 31-year publication dataset that captures the emergence of a new scientific field. The results indicate the development of a more interdisciplinary, globally connected science as opposed to science driven by single experts.

**1. Introduction**

Work dating back to the ancient Greeks argues that humanity can be seen as a complex social system or super-organism. In this perspective, people are viewed as analogous to nerve cells that are interconnected by communication channels, collectively forming a 'global brain'[1]. By adopting this philosophy one is led to believe – and hope, given the nearly constant human cognitive abilities – that there is a general trend towards the formation of a more global knowledge production and consumption dynamics exploiting the integration of social systems in concert with technological and biological systems.

This paper presents a suite of approaches and measures aimed at the quantitative study of the evolution of scientific co-authorship teams into tightly coupled global networks. The approaches and measures are applied to study the emergence of a 'global brain' in a comprehensive scholarly publication dataset witnessing the evolution of a new scientific field. In particular, we focus our attention on the study of successful co-authorship teams, where by *success* we refer to the impact of the collaboration as measured in terms of the number of publications and the relative number of citations received by each produced paper. Our aim is the discrimination between two different scenarios; a first one in which scientific progress is driven by the brilliance of a few single authors and a second one in which the collective knowledge consumption and production is mainly pushed forward by author teams and their complex set of interactions. Note that author teams refer to sets of authors that collaborated on diverse papers in different combinations.

In order to quantitatively scrutinize those two possibilities, section 2 presents a novel weighted graph representation which represents co-authorship relations as well as the number of produced papers and attracted citations. Section 3 introduces a publication dataset comprising 614 papers published in the area of information visualization by 1,036 unique authors between 1974 and 2004. The weighted graph representation is applied to encode the co-authorship relations in this dataset as well as their impact. Section 4 presents a suite of known and novel analysis and visualization techniques that can collectively be applied to study the emergence of a 'global brain'. All techniques are exemplarily applied to analyze and visualize different time slices of the 31-year publication dataset introduced in section 3. The analysis of this particular dataset confirms our hypothesis that a 'global brain' comprised of larger highly successful co-authorship teams is developing. Section 5 summarizes the contributions of this paper and outlines planned future work.

**2. Weighted Graph Representation**

An emerging goal of researchers in diverse fields of science is to improve our understanding of the topological, dynamical and organizational properties of social, biological, and technological networks[2-10]. Most of those networks do not exist in isolation, but are embedded in a network ecology. Striking examples are provided by paper networks which are created and utilized by authors, food webs that strongly depend on the environmental conditions, and protein interaction networks that are working in concert with many other biological entities. While prior work has begun to analyze and model the formation of co-authorship networks[11,12] and the major interactions among, e.g., paper and author networks[13], it is often desirable from an algorithmic point of view to deal with just one network. In general, weighted networks add an extra dimension to network representation since they can be employed to project important features of one network (e.g., the number of papers produced by a co-author team or the number of citations received by a paper) onto a second network (e.g., the network of co-authors that produced the set of papers). These so called weighted graph representations have been used to study airline connection, scientific collaboration, metabolic paths and several other networks[14,15]. Furthermore, a weighted graph can often be conveniently represented as an un-weighted multigraph and thus standard techniques for un-weighted graphs can be applied[16].

As an example, Figure 1 shows a coupled author-paper network (left) and one possible weighted graph representation of it (right). The coupled author-paper network has two types of nodes: five authors (blue circles, labeled Author_1-Author_5) and six papers (black triangles, labeled 1-6). It has three types of edges: Papers are linked via directed 'provided input to' edges (in black) that correspond to inverse 'cited by' edges. Authors are connected by undirected co-authorships edges (in blue). Light green directed edges denote the flow of information from papers to authors, and from authors to new papers via 'consumed' (or cited by) and 'produced' (or authored by) relations. Although these 'consumed' and 'produced' edges are not further discussed in this paper they are the only ones that interconnect the co-author and paper-citation networks. Arc directions denote the direction of information flow.

In the example given in Figure 1 (right), paper 1 was authored by author numbers 2, 3 and 4, paper 2 by 1, 3 and 4, paper 3 and 4 by author 2, paper 5 by author 5, and paper 6 by author 2 and 4. Hence there are 5 co-authorship edges. Author 5 has not collaborated so far. Paper 6 cites papers 1 and 2. Note that consumed papers do not need to be cited.

Figure 1 (left) shows the co-author network exclusively. The size of author nodes corresponds to the number of papers the author published. Co-authorship edges are width coded according to the number of co-authorship collaborations. Not depicted is information on the number of citations that papers received.

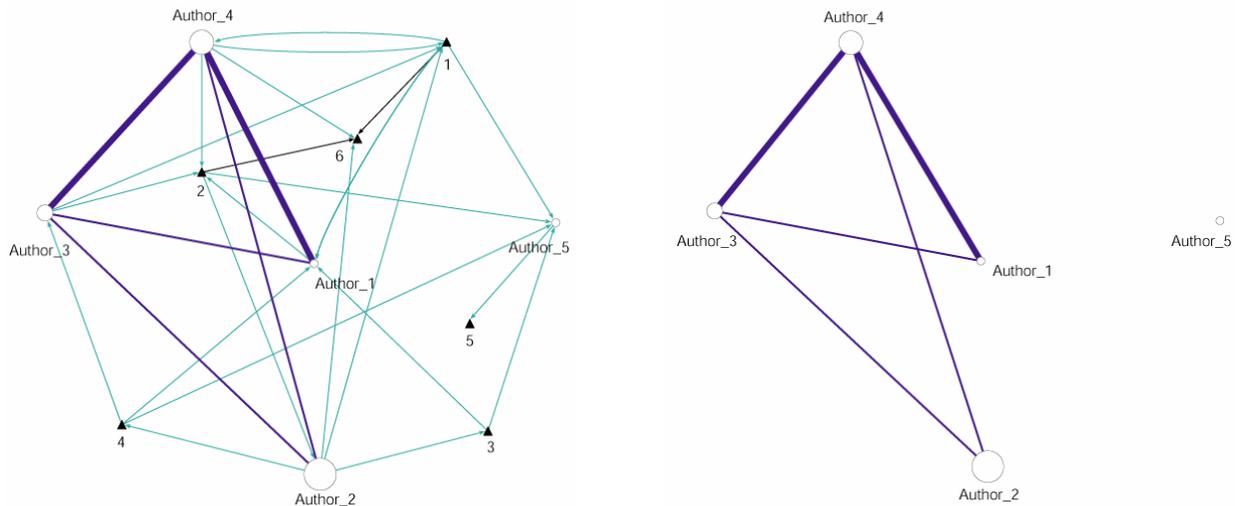

**Figure 1.** Coupled author-paper network and a weighted graph representation.

While the number of papers co-authored by a set of authors might well be an indicator of the impact of a co-authorship team, we are interested in studying the impact of co-authorship teams based on the citations that co-authored papers received. Therefore, we define an *impact weight* via the sum of the *normalized impact* of the paper(s) that resulted from a co-authorship.

To do so, we make the subsequent assumptions:
- The existence of a paper $p$ is denoted with a unitary weight of 1, representing the production of the paper itself. This way, papers that do not receive any citations do not completely disappear from the network.
- The *impact* of a paper grows linearly with the number of citations $c_p$ the paper receives.
- Single author papers do not contribute to the co-authorship network weight or topology.
- The impact generated by a paper is equally shared among all co-authors.[1]

The *impact* of a co-authorship edge can now be defined as the number of papers that were jointly co-authored plus the number of citations acquired by these papers. Formally, the *impact weight* $w_{ij}$ associated with an edge *(i,j)* is defined as

$$w_{ij} = \sum_p \frac{(1+c_p)}{n_p(n_p-1)},$$

where the index $p$ runs over all papers co-authored by the authors *i* and *j*, and $n_p$ is the number of authors and $c_p$ the number of citations of paper $p$, respectively. The normalization factor $n_p(n_p-1)$ ensures that the sum over all the edge weights per author equals the number of citations divided by the number of authors. It is worth remarking that since we are interested in the role of authors and their interaction in the co-authorship networks, only the co-authored papers are considered in the weights calculation. In addition, summing up all the weights in a co-authorship network that were contributed by a paper results in the impact of this paper. For $c_p=0$, the weight $w_{ij}$ is equal to the number of papers normalized by the number of authors. It is important to stress that in our assumption the weights are always symmetric $w_{ij}= w_{ji}$.

---
[1] Note that this might not accurately reflect reality as different authors typically contribute different skills and amounts of work to a paper. If data on the specific author contributions is available this assumption should be reconsidered.

To give an example, for a paper with two authors, the complete impact (1 plus all received citations) goes to one co-authorship edge. For a paper with 3 authors, the impact value is divided by three and added to the weights of three co-authorship edges. For example, Figure 2 (left) shows a scenario where three authors produce a paper that receives two citations. Adding up the two weights ($w_{ij}+ w_{ji}$) per edge for all edges results in 3 and hence equals the number of citations plus one. Summing up the weights per author results in a weight of one and summing up the weights for all authors results again in a weight of 3. If a paper has four authors, the impact needs to be divided by 6. Figure 2 (right) shows the weights added by a paper that did not receive any citations yet and was authored by four people. All arrow weights are 1/12, times 12 arrows results in the unitary weight of 1, representing the production of the paper itself. Summing up the contributions per author results in 1/4, which multiplied by the number of all authors again results in 1.

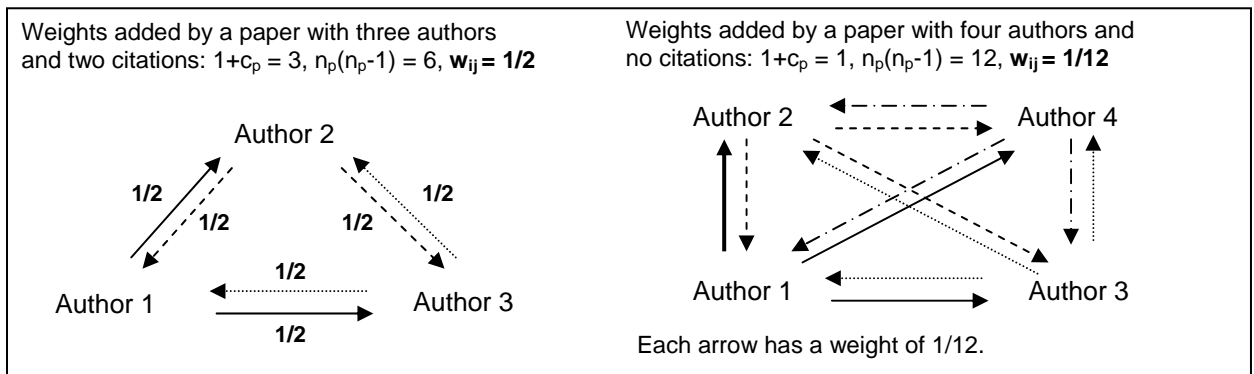

**Figure 2.** Exemplification of the impact weight definition.

It is worth remarking that in previous complex networks studies[17] a different weight has been used to weight co-authorship networks edges. These studies focused on an evaluation of the strength of the connection in terms of the continuity and time share of the collaboration. Our focus is on the productivity (number of papers) and the impact (number of papers and citations) of the collaboration. Moreover, prior work in information science[18] has considered and contrasted three different approaches to allocating citational credit: only the first author receives credit, each author receives full credit, and fractions are awarded to each co-author. To our knowledge, we are the first to suggest that citational credit should be awarded to co-author relations so that the collective success of co-authorship teams – as opposed to the success of single authors – can be studied.

## 3. The InfoVis Contest Dataset

To study the structure and evolution of successful co-author teams we use the InfoVis Contest 2004 dataset.[2] This dataset is unique in that it documents the birth and growth of a new field of science – information visualization – between 1974 and 2004. The dataset was made available for the InfoVis Contest in 2004 (http://www.cs.umd.edu/hcil/iv04contest/). It contains all papers from the ACM library that are related to information visualization research.

From the three major InfoVis Conferences, the ACM library does capture the IEEE InfoVis Conference papers, but fails to cover papers presented at the annual InfoVis Conference in London or the annual SPIE Visualization and Data Analysis Conference in San Jose. It also does not contain papers from the new *Information Visualization* journal or books. Hence, only a partial picture of the domain can be drawn.

Summary statistics for the dataset are given in Figure 3 (left). Exactly 614 papers were published by 1,036 unique authors between 1974 and 2004. Interestingly, the ACM library does cover a steadily increasing amount of InfoVis papers from 1974-1996. However, the number of papers captured (or retrieved) in later years, 1997-2004, is steadily decreasing which might be due to alternative publication venues. For each paper published in a given year, the number of references made in the papers to older publications is given as well as the total number of citation counts received by papers published in this year. Younger papers obviously had less time to receive citations. Only within-set paper citations are available to characterize the impact of co-authorships. Citations to papers within the dataset from papers outside of the dataset will be missed. Future work will apply the proposed weighted graph representation and measurement set to larger publication datasets to address this common data coverage issue.

---
[2] The dataset is documented and available online at http://iv.slis.indiana.edu/ref/iv04contest/.

Figure 3 (left) also gives information on the number of new unique authors that enter the field each year. Figure 3 (right) provides more details on the size of co-author teams over time. Please note that the dataset was pruned and only statistically significant years in which more than 10 papers were published, i.e., 1988-2002, are shown. Author team sizes that occurred only once are not graphed as, for instance, the 11 and 12 author teams that published in 1995. Values are normalized by total number of articles every year.

There seems to be a decrease in the number of papers published by a single author confirming research by Beaver and Rosen[19] on the increase in the number of authors per paper as a field matures and institutionalizes. However, our initial question: "Is scientific progress driven by the brilliance of a few single authors or by effectively collaborating author teams" cannot be answered.

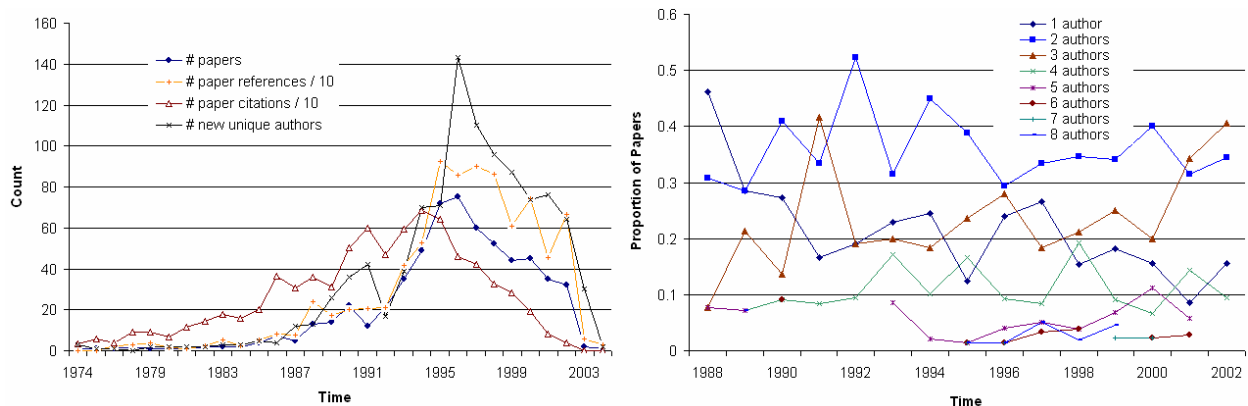

**Figure 3.** Number of papers, citation and reference counts, and new unique authors over the 31-year time frame (left) and proportion of the number of authors per paper (right).

### 4. Weighted Network Analysis & Visualization

This section presents a suite of known and novel measures that can collectively be applied to study the impact of co-authorship teams on a local and global scale. The section starts with a visualization of three time slices of the weighted co-author network to give a birds eye view of its structure and evolution over time. Subsequently, we introduce measures (1) to identify highly productive and influential authors, (2) to analyze co-author degree and strength distributions, (3) to examine the size and distribution of connected components, and (4) to determine the homogeneity of co-authorship weights per author.

All measures are exemplified using different subsets of the 31-year citation dataset:
- Three times slices are used for the visualization presented in section 4.1 and measures given in 4.4: papers published from 1974-1984, 1974-1994 and 1974-2004. Although the 1974-1994 data dominates the 1974-1999, and 1974-2004 data, the alternative choice, i.e., using papers published in or before 1984, in 1985-1999, and 2000-2004, would not work as the time span is simply too short for the evolution of meaningful co-authorship networks.
- The latter two time slices 1974-1994 and 1974-2004 are also used in section 4.2, 4.3., and 4.4.
- The complete 1974-2004 dataset is used for measures introduced in section 4.2.

#### *4.1 Visualization of Network Evolution*
To gain a better understanding of the structure and dynamics of co-authorship relations, we plotted the co-author network for three different time slices. In particular, the Kamada and Kawai[20] layout algorithm in the Pajek visualization toolkit[21] was applied to layout the weighted co-author networks for three time slices: 1974-1984, 1974-1994 and 1974-2004. Nodes are used to represent authors, edges denote co-authorship relations. The node area size reflects the number of single-author and co-authored papers published in the respective time period. Node color indicates the cumulative number of citations received by an author. Hence, large, black-shaded author nodes denote authors that published many highly cited papers. Edge color reflects the year in which the co-authorship was started. Edge width corresponds to the impact weight introduced in the previous section. Thus thick, black-shaded co-authorship edges represent multiple successful co-authorships. Color and size coding of nodes and edges are identical for all three figures as is the placement of author nodes.

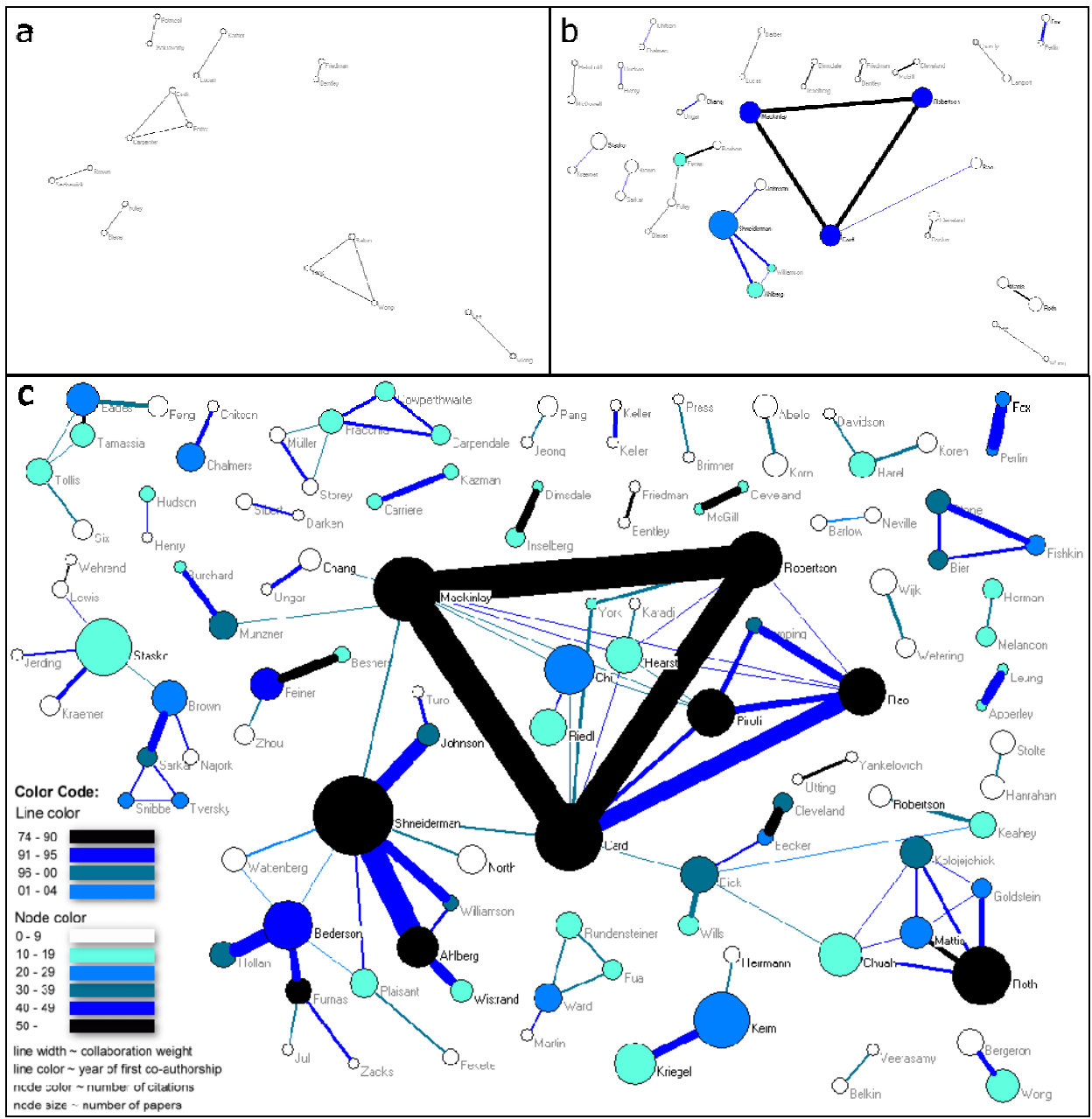

**Figure 4.** Weighted co-author network for papers published in 74-84 (a), 74-94 (b) and 74-04 (c). The color legend applies to all three graphs.

Figure 4a shows all authors that published in or before 1984 as well as their co-authorship relations. Figure 4b shows all authors from the 74-94 network that are connected by edges which have a impact weight of equal or more than 1.0. The 74-04 co-author network in Figure 4c shows all authors that are connected by edges which have an impact weight of equal or more than 2.0. That is, a threshold value was applied to the last two time slices to preserve the highest impact edges and to ensure the readability of the graph.

Interestingly, researchers working at research labs show very strong co-authorship relations. For example, Mackinlay, Robertson and Card (all three worked at Xerox Parc until 1996 when Robertson joined Microsoft Research) co-authored 12 papers, six of which were co-authored by all three. The 12 papers received 202 citations from papers in the selected dataset. The most highly cited paper is "Cone Trees: animated 3D visualizations of hierarchical information" with 70 citations. The Rao, Pirolli and Card co-authorship triangle is based on nine papers. Each of those papers had two authors and all nine papers attracted 87 citations in total.

Researchers working at Universities, e.g., Shneiderman, show a very different co-authorship pattern in which one high weight node (a professor) is well connected to a large number of lower weight nodes (a steady stream of coming and going students). Shneiderman, Ahlberg and Williamson co-authored five papers, where only one was co-authored by all three of them. These five papers were cited 67 times.

The three time slices show the emergence of a new field of science that grows out of many unconnected or weakly connected (low weight) nodes, merging into multiple connected clusters and finally in an almost fully connected network that essentially accumulates the largest weights. This illustrates the formation of a complex collaboration network where the production of scientific knowledge is driven by the global web of interaction and the emerging paths of communication and collaboration. A note of caution in this interpretation is necessary because of the limited size of the data set. In larger research communities more than one cluster at a time could appear. A more definitive assessment of the present results should be provided for extensive datasets in homogeneous disciplines and subsequently in their global interconnections.

### *4.2 Identification of Highly Productive and Influential Authors*

Subsequently, we are interested in identifying the most highly productive and influential authors. Four measures are introduced and contrasted: node degree, citation strength, productivity strength, and betweenness[22].

The *degree k* of a node, also called point centrality, equals the number of edges attached to the node. In our case the degree of an author refers to the number of unique co-authors an author has acquired.

A second centrality measure, the *citation strength $S_c$* of a node *i* is defined as

$$s_c(i) = \sum_j w_{ij} .$$

The citation strength is growing with the number of papers and citations, providing a quantitative evaluation of the total impact of an author. We obtain a third measure, if we count the number of papers an author team produced but disregard the citations these papers attracted, i.e., $c_p=0$. We call this measure *productivity strength $s_p(i) = s_c(i)|_{c_p=0}$*. It is worth remarking that both strength definitions considers only co-authored papers since the weights are defined only by the interaction network of coauthored papers.

The fourth measure, the *betweenness* of a node *i*, is defined to be the fraction of shortest paths between pairs of nodes in the network that pass through *i*, i.e., the betweenness denotes the extent to which a node (author) lies on the paths between other authors.

The top ten authors and their values for all four measures for the complete 1974-2004 time span are given in Table 1. While several authors are listed in more than one top ten list, their exact placement is rather different. Ben Shneiderman at the University of Maryland appears to be the clear winner according to three of the four measures. Interestingly, the top three author sets are identical for all four measures. However, there are many names that appear in only one top ten list, reflecting the fact that the measures clearly evaluate different qualities of an author. While degree refers to the raw number of collaborators, productivity strength detects authors with the largest number of coauthored papers since the strength is only accounting for coauthored papers. The citation strength, instead, focuses on the number of citations received by the author on coauthored papers, and finally the betweeness selects authors that are very important in connecting different author group and thus act as communication switch. Authors rank differently according to each measure, showing a rich ecology in which each author has a different role and characteristic. Note that a close examination of Figure 4 allows one to confirm the data displayed in Table1.

**Table 1.** Author ranking based on degree (# co-authors), productivity strength (# produced papers), citation strength (# received citations), and betweenness (# of shortest paths that pass through this author).

| Degree k | # | Productivity Strength $S_p$ | # | Citation Strength $S_c$ | # | Betweenness | # |
|---|---|---|---|---|---|---|---|
| B._Shneiderman | 23 | B._Shneiderman | 7.62 | S._K._Card | 88 | B._Shneiderman | 10893 |
| J._D._Mackinlay | 17 | S._K._Card | 5.71 | J._D._Mackinlay | 67 | S._K._Card | 10618 |
| S._K._Card | 17 | J._D._Mackinlay | 4.37 | B._Shneiderman | 66 | J._D._Mackinlay | 8357 |
| G._Robertson | 16 | Daniel_A._Keim | 4.11 | G._Robertson | 64 | Stephen_G._Eick | 7420 |
| Allison_Woodruff | 15 | Steven F. Roth | 3.96 | Christopher Ahlberg | 36 | Chris_Olston | 5165 |
| Lucy_T._Nowell | 15 | John_T._Stasko | 3.92 | R._Rao | 34 | Ben_Bederson | 4791 |
| Roberto_Tamassia | 15 | Stephen_G._Eick | 3.67 | Ben_Bederson | 25 | Mei_C._Chuah | 4718 |
| Ben_Bederson | 15 | G._Robertson | 3.46 | Peter_Pirolli | 21 | G._Robertson | 3187 |
| Harpreet_S._Sawhney | 14 | Ben_Bederson | 3.40 | Steven_F._Roth | 20 | Steven_F._Roth | 2063 |
| M._Stonebraker | 14 | Marc_H._Brown | 3.33 | Brian_Johnson | 17 | E._H._-H._Chi | 1718 |

### *4.3 Analysis of Co-Author Degree and Strength Distributions*

While the measures discussed in subsection 4.1 provide information about single authors, they fail to capture emergent global network properties. Therefore, an analysis of co-author degree and strength distributions was conducted to analyze the size of the most successful co-author teams over time. Figure 5 shows the resulting distributions for the 1974-1994 and the 1974-2004 time slices. In each graph the cumulative distribution $P_c(x)$ is the probability that any given author has a centrality value larger than a given value *x*.

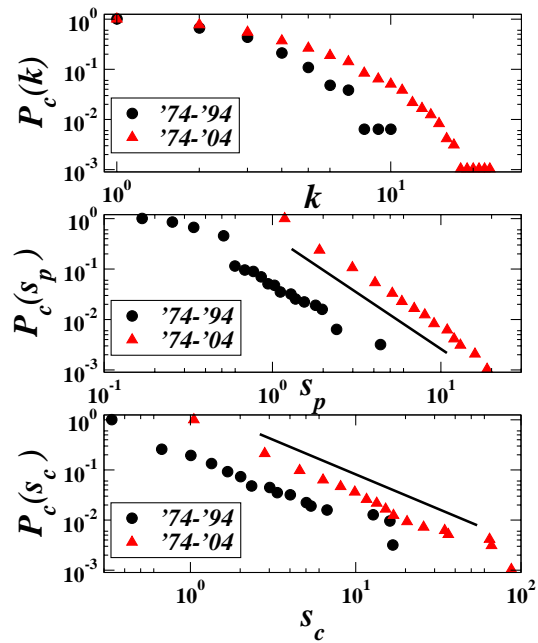

**Figure 5.** Co-author degree (top), productivity strength (center) and citation strength distributions (bottom) in two different time slices of the InfoVis dataset. For the productivity and citation strength distributions, the solid line is a reference to the eye corresponding to a heavy-tail with power-law behavior $P(x) = x^{-\gamma}$ with $\gamma = 2.0$ (top right) and 1.4 (bottom left), respectively.

As can be seen, the distributions are progressively broadening in time, developing heavy tails. This implies that a high variability of connectivity and impact is setting up in the system. The heavy tails indicate that we are moving from a situation with a few authors of large impact and a majority of peripheral authors to a scenario in which impact is spread over a wide range of values with large fluctuations for the distribution. Indeed, for such a distribution the variance is virtually unbounded indicating extremely large statistical fluctuations. In other words, if one picks up an author at random then it is very likely that we find values of the various measures which are far from the average.

Figure 6 plots the publication strength $S_p$ and the citation strength $S_c$ of authors versus the degree of authors (number of co-authors) for the 74-04 time slice. There are two definite slopes: one for low degrees and another one for high degrees (much faster). This implies that the impact and productivity grows faster for authors with a large number of co-authorships as if a sort of cooperative beneficial effect is reached by larger collaboration groups. For the citation impact graph we see that a few very high degree nodes receive an disproportionately high amount of citations. Closer examination of the data reveals that the three outliers are S. K. Card, J. D. Mackinlay, and B. Shneiderman.

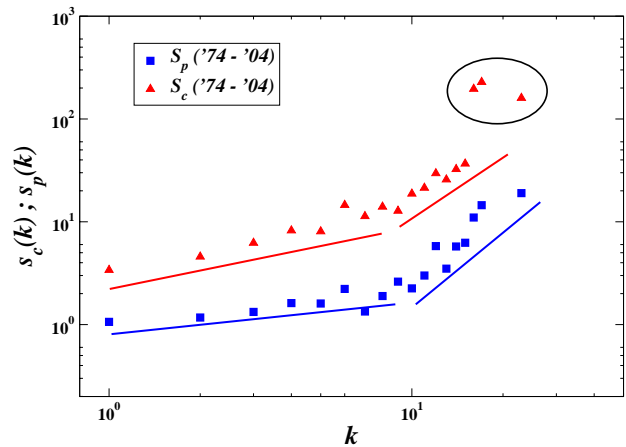

**Figure 6.** Publication strength $s_p$ and citation strength $s_c$ of authors over the degree of authors. Solid lines are a guide to the eye indicating the presence of two different regimes as a function of the co-authorship degree $k$.

### *4.4 Examining the Size and Distribution of Connected Components*

Next, we measured the relative size of the largest connected component (as the percentage of authors in the graph contained in this component) and of all the other connected components for three time slices. The size of the component is calculated in four different ways: $G_N$ is the relative size measured as the percentage of nodes within the largest component. $E_g$ is the relative size in terms of edges. $G_{sp}$ is the size measured by the total strength in papers of authors in the largest component. Finally, $G_{sc}$ is the size measured by the relative strength in citations of the authors contained in the largest component. Table 2 shows the steady increase of the giant component in terms of all four measures for the 1974-1994, 1974-1999, and 1974-2004 time slices. It is interesting to note that while the largest component amounts to 20% of the total authors of the network, it accumulates more than 40% of the citation impact. This implies that the emerging connected cluster of authors is able to disproportionately produce more papers and to attract more citations.

**Table 2.** Relative size of the giant component in terms of the four measures as function of time.

|  | **1974-1994** | **1974-1999** | **1974-2004** |
|---|---|---|---|
| **$G_N$** | 8.30% | 12.50% | 15.50% |
| **$E_g$** | 14.40% | 16.50% | 20.20% |
| **$G_{sp}$** | 10.10% | 21.80% | 24.10% |
| **$G_{sc}$** | 19.30% | 38.80% | 40.60% |

In order to provide a more thorough representation of the clustering dynamics we show the zipf plot of the relative sizes of the graph components for the 1974-1994 and the 1974-2004 time slices. The zipf plot is obtained by ranking all components of the co-authorship graphs in decreasing order of size and then plotting the size and the corresponding rank of each cluster on a double logarithmic scale. In Figure 7 we show that the largest component is steadily increasing both in size and impact. In fact, all four curves cross, indicating that in general the few best ranked components increase at the expense of the smaller components (clusters) of authors. It is also interesting to note that the second largest component is much smaller than the largest one, pointing to the formation of one dominating cluster of authors which publishes a large number of highly cited papers.

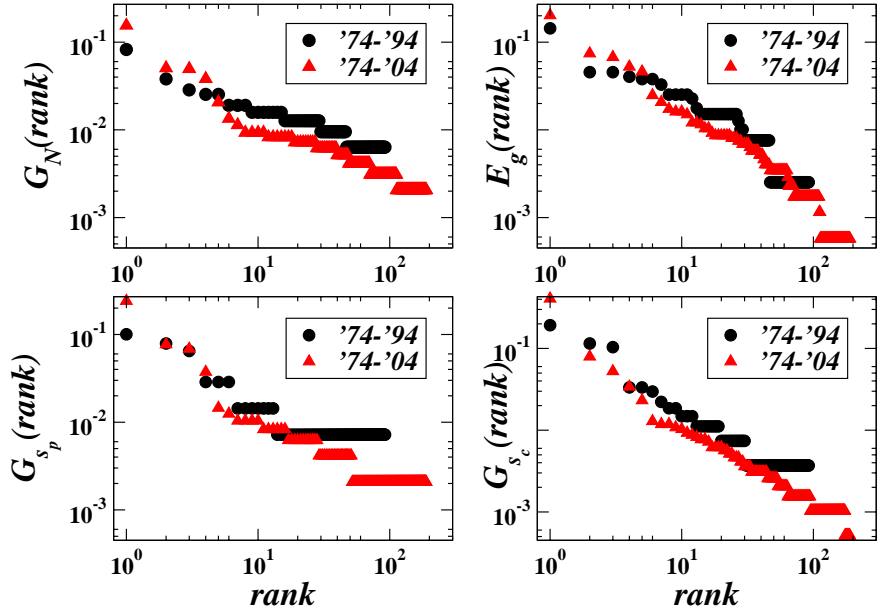

**Figure 7.** Zipf plot of various attributes of the network connected components. Relative size of authors (top left), co-authorship edges (top right), publication strength (bottom left) and citation strength (bottom right).

### *4.5 Determining the Homogeneity of Co-Authorship Weights per Author*

Another important issue concerns the inspection of the homogeneity of co-authorship weights per author. We are interested to find out if the impact of an author is spread evenly over all her/his co-authors or if there exist 'high impact co-authorship edges' that act as strong communication channels and high impact collaborations. In particular, we would like to know if the collaboration of authors leads to an impact that goes beyond the sum of the impact of single authors. To answer this question, we apply a novel local entropy-like measure defined as

$$H_{Sx}(i) = \frac{-1}{\log k_i} \sum_j \left(\frac{w_{ij}}{s_x(i)}\right) \log\left(\frac{w_{ij}}{s_x(i)}\right),$$

where *x* can be replaced by *p* or *c* denoting the paper strength or citation strength respectively, $k$ is the degree of node *i* and $w_{ij}$ is the impact weight.

This quantity is bounded by definition between $0 \leq H_i \leq 1$. It measures the level of disorder with which the weights are distributed in the neighborhood of each author. The homogeneous situation corresponds to having all weights equal, i.e., $w_{ij}=w$ and $s_i=k_i w$. In this case the entropy is $H=1$. The opposite case is when one single connection accumulates a disproportionate weight at the expenses of all others; i.e. $H \to 0$. In Figure 8 we report the entropy spectrum $H(k)$ that provides the average entropy for authors with degree $k$, and allows the discrimination of the impact distribution for authors with different numbers of co-authorships.

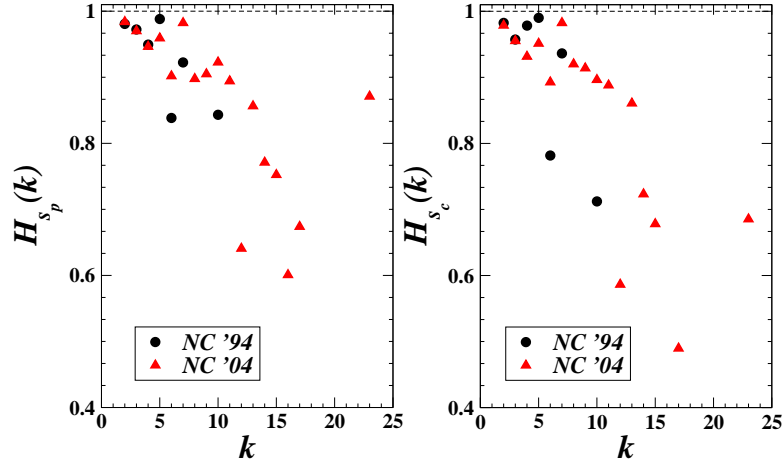

**Figure 8.** Entropy spectrum for the productivity strength $S_p$ (left) and citation strength $S_c$ (right) weighted graph.

It is important to note that for all large values of $k$ the entropy is decreasing. This signals that highly connected authors develop a few collaborations that have a very high strength compared to all other edges. These highly productive and high impact co-authorship edges are not homogeneously distributed – a few "synaptic" connections in the global brain are more active and successful than most others. This is also consistent with the distribution of impact weights that exhibit a power law behavior signaling the high heterogeneity of co-authorship impact in the entire system.

Finally, we report in Figure 9 the scatter plot of the citation strength due to collaborations $S_c$ and the citation strength due to single authored papers $S_{cs}$ of each author for the 1974-2004 time slice. The single author citation strength is simply obtained by the sum of $(1+c_p)$ of single authored papers. Note that the plot reports only authors with both co-authored and single authored papers. It is important to remark that the total impact generated by single author papers is just a few percent of that generated by collaborations and it is decreasing along the years. As expected, it is possible to find a correlation between the two impacts that, however, is not extremely strong. In addition, a best fit provides a functional behavior $S_c \sim S_{cs}^{0.75}$.

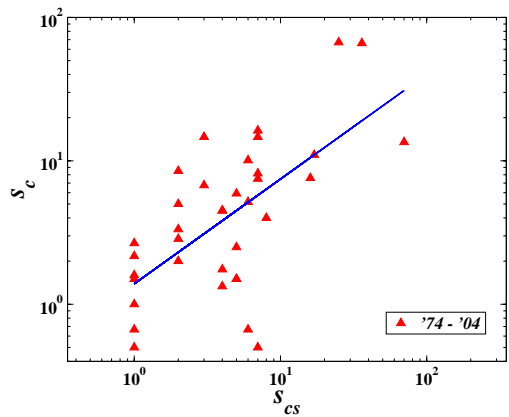

**Figure 9.** Scatter plot of citation strength due to collaborations $S_c$ and due to single authored papers $S_{cs}$

This implies that the impact of co-author teams is not linearly related to the impact of a single author. The sub-linear behavior is most likely due to the fact that authors which are able to produce higher impact single authored papers (e.g., professors) also collaborate with lower impact authors (e.g., students) resulting in a slower increase of the collaboration impact. Hence, while it might be beneficial for high impact authors to publish single author papers, the selfish perspective does not seem to prevail on the cooperative tendency that leads to the development of an increasingly larger co-authorship network.

## 5. Summary and Future Work

This paper presented a novel weighted graph representation of coupled networks that can be applied to study the local and global properties of co-authorship networks and to dynamically visualize the changing impact of co-authorship relations. Based on the weighted graph representation we defined diverse measures to determine and contrast the centrality and impact of single authors based on the following measures:

- global co-author degree and strength distributions,
- the size of the largest connected component and its growth over time,
- the homogeneity of impact weights per author using a novel entropy-like measure, and
- the citation strength due to co-authored and single author papers.

While it is clear that the scientific impact and influence of a collaboration cannot be measured in terms of papers and citations alone, the developed graph representation and measures allow one to objectively define and measure the impact of co-authorship teams as a partial indicator of their success.

The weighted graph representation and measures were applied to analyze a 31-year publication dataset that captures the emergence of a new field of science. Major results comprise the identification of key authors, a first glimpse of the richness of the co-author ecology in which each author has a different role and characteristics, a change from a situation in which very few authors have high impact to a scenario in which impact is spread over a wide range of impact values, a steady increase of the size of the largest connected component, and an inhomogeneous distribution of high impact edges per author. All these results obtain for this particular dataset point towards the emergence of a 'global brain', i.e., a more interdisciplinary, globally connected science as opposed to science driven by single experts. This is very good news indeed as science driven by single authors will not scale to process, understand and manage the amounts of information and knowledge available today. However, science driven by high impact co-authorship teams will be able to dynamically respond to the increasing demands on information processing and knowledge management.

Future work will apply the proposed weighted graph representation and measurement set to diverse publication datasets. The normalization of citations for young papers is a serious issue for the analysis of recent developments. Issues due to the 'cut' of a dataset (e.g., papers in the very first few years of a dataset receive very few citation edges from papers within the set; citations from outside papers need to be accounted for) need more thorough examination. A closer mathematical and empirical examination of the correlation among the four centrality and impact measures of authors and their relation to prior work in bibliometrics[23,24] is expected to lead to new insights into the co-authorship dynamics. Last but not least, it will be interesting to study the utility of the proposed graph representation and measures to develop more robust and scalable methods for network pruning and visualization.

**Acknowledgements**

We would like to thank Steven A. Morris and the anonymous reviewers for detailed comments on a previous version of this paper and Blaise Cronin for pointing out related work. We appreciate the effort by Jean-Daniel Fekete, Georges Grinstein, Catherine Plaisant and others in providing the InfoVis Contest 2004 dataset. This work is supported by a National Science Foundation CAREER Grant under IIS-0238261 to the first author.

**References**

1. Bloom, H. Global Brain: The Evolution of Mass Mind from the Big Bang to the 21st Century. John Wiley & Sons, Inc.: New York, 2000.
2. Barabási, A.-L. Linked. Perseus Books Group, 2002.
3. Barabási, A.-L.; Albert, R. Emergence of scaling in random networks. Science 1999, 286, 509-512.
4. Wasserman, S.; Faust, K. Social Network Analysis : Methods and Applications (Structural Analysis in the Social Sciences, 8). Cambridge University Press, 1994.
5. Pastor-Satorras, R.; Vespignani, A. Evolution and structure of the Internet. Cambridge University Press: Cambridge, 2004.
6. Börner, K.; Chen, C.; Boyack, K. Visualizing Knowledge Domains. In Annual Review of Information Science & Technology. Cronin, B., Ed.; Information Today, Inc./American Society for Information Science and Technology: Medford, NJ, 2003, p 179-255.
7. Newman, M. E. J. Scientific collaboration networks. II. Shortest paths, weighted networks, and centrality. Physical Review E 2001, 64, 016132.
8. Newman, M. E. J. Scientific collaboration networks. I. Network construction and fundamental results. Physical Review E 2001, 64, 016131.
9. Dorogovstev, S. N.; Mendes, J. F. F. Evolution of Networks. Oxford University Press, 2003.
10. Amaral, L. A. N.; Scala, A.; Barthelemy, M.; Stanley, H. E. Classes of small-world networks. Proc. Natl. Acad. Sci. USA 2000, 97, 11149-11152.
11. Ramasco, J. J.; Dorogovtsev, S. N.; Pastor-Satorras, R. Self-Organization of collaboration networks. Physical Review E 2004, 70, 036106.

12. Guimera, R.; Uzzi, B.; Spiro, J.; Amaral, L. A. N. Team assembly mechanisms determine collaboration network structure and team performance. Northwestern University preprint 2004.
13. Börner, K.; Maru, J.; Goldstone, R. The Simultaneous Evolution of Author and Paper Networks. Proceedings of the National Academy of Sciences USA 2004, 101(Suppl_1), 5266-5273.
14. Almaas, E.; Kovacs, B.; Viscek, T.; Oltvai, Z. N.; Barabasi, A.-L. Global organization of metabolic fluxes in the bacterium, Escherichia coli. Nature 2004, 427.
15. Barrat, A.; Barthelemy, M.; Pastor-Satorras, R.; Vespignani, A. The architecture of complex weighted networks. Proc. Natl. Acad.. Sci. USA 2004, 101, 3747-3752.
16. Newman, M. E. J. Analysis of weighted networks. cond-mat/0407503 2004.
17. Newman, M. E. J. Coauthorship networks and patterns of scientific collaboration. Proceedings of the National Academy of Sciences USA 2004, 101, 5200-5205.
18. Cronin, B.; Overfelt, K. Citation-Based Auditing of Academic Performance. Journal of the American Society for Information Science 1994, 45, 61-72.
19. Beaver, D.; Rosen, R. Studies in scientific collaboration, Part I: The professional origins of scientific co-authorship. Scientometrics 1978, 1, 65-84.
20. Kamada, T.; Kawai, S. An algorithm for drawing general undirected graphs. Information Processing Letters 1989, 31, 7-15.
21. Batagelj, V.; Mrvar, A. Pajek: A Program for Large Network Analysis. Connections 1998, 21, 47-57, (Available at http://vlado.fmf.uni-lj.si/pub/networks/pajek/).
22. Freeman, L. C. A set of measures of centrality based upon betweenness. Sociometry 1977, 40, 35-41.
23. Crane, D. Invisible Colleges: Diffusion of Knowledge in Scientific Communities. The University of Chicago Press: Chicago, 1972.
24. White, H. D. Author-centered bibliometrics through CAMEOs: Characterizations automatically made and edited online. Scientometrics 2001, 51, 607-637.